\documentclass{epl}

\title{Non-equilibrium hydrodynamics of a rotating filament}
\author{H. Wada \and R. R. Netz}

\institute{                    
 Physik Department, Technical University Munich, 85748 Garching, Germany
}
\pacs{87.15.-v}{Biomolecules: structure and physical properties}
\pacs{47.20.Ky}{Nonlinearity, bifurcation, and symmetry breaking}
\pacs{47.63.mf}{Low-Reynolds-number motion}

\newcommand{\bra}{\langle}
\newcommand{\ket}{\rangle}
\newcommand{\vecOm}{\boldsymbol{\Omega}}
\newcommand{\vecmu}{\boldsymbol{\mu}}
\newcommand{\vecxi}{\boldsymbol{\xi}}

\begin{document}

\maketitle

\begin{abstract}
The nonlinear dynamics of an elastic filament that is forced to rotate 
at its base is studied by hydrodynamic simulation techniques;
coupling between stretch, bend, twist elasticity and thermal fluctuations
is included.
The twirling-overwhirling transition is located and found to be strongly discontinuous.
For finite bend and twist persistence length,
thermal fluctuations lower the threshold rotational frequency,
for infinite persistence length the threshold agrees with previous analytical
predictions.
\end{abstract}

The dynamics and morphology of elastic filaments are of interest
in various fields encompassing systems on many different 
length scales~\cite{goriely}.
Relevant examples are provided by biopolymes and bio-assemblies, 
including DNA, actin filaments, microtubules, and
multicellular organisms like {\it Bacillus subtilis}~\cite{wiggins}.
Modern micro-manipulation techniques allow to
observe and analyze filament dynamics on the single-molecule level~\cite{kaes},
prompting a detailed  understanding of the combined
effects of fluctuations, hydrodynamics and elasticity.
Static properties of elastic rods under external force loads 
have a long history of study, dating back to Euler~\cite{landau}.
Of current interest is the time-dependent  
behavior of such driven filaments and the possibility
of shape bifurcations, especially on the biologically relevant
nano-to-micro length scales,
where viscous hydrodynamic dissipation dominates over inertia.

In this Letter, we describe a hydrodynamic simulation technique 
for elastic filaments with arbitrary shape and rigidity
and subject to external forces or boundary conditions,
including full coupling between thermal, elastic and hydrodynamic
forces in low-Reynolds-number flow.
We apply our method to the whirling dynamics
of a slender filament that is axially rotated at one end at frequency $\omega$ 
(while the other end is free).
This model system has first been studied analytically by Wolgemuth
{\it et al.}~\cite{wolgemuth}, who showed that a critical frequency $\omega_c$ 
separates whirling (steady-state crankshafting motion with
axial spinning) from twirling (diffusion-dominated simple axial
rotation) by a supercritical Hopf bifurcation (i.e., a continuous 
shape transition).
Subsequently, a zero-temperature 
simulation has been performed using the immersed boundary method~\cite{lim},
where the microscopic structure of the filament was
modelled by interconnected springs. In contrast to
analytic predictions, 
the filament was shown to undergo a subcritical (i.e. discontinuous) 
shape transition from twirling to a strongly bent state 
where the filament almost folds back on itself (termed {\it overwhirling}).
The transition frequency $\omega_c$  was found to
be smaller by a factor of 3.5 compared to the analytic estimate.
Using our simulation technique, we first map out  the stability diagram 
in the absence of thermal fluctuations
as a function of the driving rotational
frequency; $\omega_c$ is identified as the
instability point of twirling and
agrees quantitatively with analytical prediction\cite{wolgemuth};
however, a stable small-amplitude whirling
state is absent and instead a  strongly bent overwhirling state
is found for all frequencies $\omega$ larger than $\omega_c$.
In the presence of thermal fluctuations the transition
frequency $\omega^*$ decreases as the filament becomes
more flexible and the discontinuous nature of the
transitions is manifested by pronounced hysteresis.

To proceed,
consider a filament with circular cross section and contour length
$L$, parameterized by arclength $s$.
A generalized Frenet orthonormal basis 
$\{\hat{\bf e}_1,\hat{\bf e}_2,\hat{\bf e}_3\}$
is defined at each point of the filament centerline ${\bf r}(s)$, 
where $\hat{\bf e}_3$ points along the tangent and  
$\hat{\bf e}_1$, $\hat{\bf e}_2$ lie in the plane normal to $\hat{\bf e}_3$ 
such that the basis forms a right-handed triad, i.e.  
$\hat{\bf e}_2=\hat{\bf e}_3\times\hat{\bf e}_1$.
The strain vector $\vecOm(s)=(\Omega_1,\Omega_2,\Omega_3)$
characterizes the shape of the filament through the 
kinematic relation 
$\partial_s\hat{\bf e}_j=\vecOm\times\hat{\bf e}_j$,
where $\kappa=(\Omega_1^2+\Omega_2^2)^{1/2}$ is the curvature
and $\Omega_3$ the twist density.
The elastic energy of an elastic  inextensible rod reads~\cite{landau}
\begin{equation}
 E = \frac{A}{2}\int_0^L ds[(\Delta\Omega_1)^2+
	(\Delta\Omega_2)^2]+
 \frac{C}{2}\int_0^Lds (\Delta\Omega_3)^2,
 \label{eq1}
\end{equation}
where $A$ and $C$ are bend and twist rigidity constants,
and $\Delta\vecOm=\vecOm-\vecOm^0$.
The ground state shape of the filament is thus controlled by the  
intrinsic curvatures $\Omega_1^0$, $\Omega_2^0$ 
and the intrinsic twist $\Omega_3^0$.  According to the
principle of virtual work, the force per length 
acting on the filament is obtained via ${\bf f}=-\delta E/\delta{\bf r}$
keeping the rotation angle about the tangent zero, i.e.,
$\delta\phi=\hat{\bf e}_2\cdot\delta\hat{\bf e}_1=0$~\cite{goldstein}.
Likewise the torque about the tangent is obtained via $m=-\delta E/\delta\phi$
without moving the centerline, i.e., $\delta{\bf r}=0$.
In the low-Reynolds-number limit, the local force and torque balance with
the viscous drag (long-range hydrodynamic interaction will be considered
later), leading to the equations $\dot{\bf r}=\mu{\bf f}$ and 
$\dot{\phi}=\mu_r m$, 
where $\dot{\bf r}=\partial{\bf r}/\partial t$ and
$\mu$, $\mu_r$ are the translational and rotational 
mobility constants.

For an efficient simulation code, it is crucial to 
introduce a laboratory-fixed reference frame 
$\{\hat{\bf x},\hat{\bf y},\hat{\bf z}\}$ and to parameterize $\vecOm$
(thereby the energy $E$) in terms of ${\bf r}$~\cite{allison}.
The unit vector
$\hat{\bf n}={{\bf t}\times\hat{\bf z}}/|{{\bf t}\times\hat{\bf z}}|$
is by construction normal to the tangent ${\bf t}\equiv\hat{\bf e}_3$.
Taking the binormal ${\bf b}={\bf t}\times{\bf n}$, we find a local 
right-handed triad $\{{\bf n},{\bf b},{\bf t}\}$ at the filament centerline
position ${\bf r}(s)$.
Since $(\hat{\bf e}_1,\hat{\bf e}_2)$ and
$({\bf n},{\bf b})$ are coplanar by definition, they are related as
$\hat{\bf e}_1+i\hat{\bf e}_2=\exp[-i\theta(s)]({\bf n}+i{\bf b})$.
The rotation angle about the tangent thus satisfies
$\delta\phi = \delta\theta+{\bf b}\cdot\delta{\bf n}$,
indicating that ${\bf r}$ and $\theta$ are not independent in 
computing the translational force ${\bf f}$.
Using the kinematic relation 
$\Omega_1=-\hat{\bf e}_2\cdot\partial_s\hat{\bf e}_3$,
$\Omega_2=\hat{\bf e}_1\cdot\partial_s\hat{\bf e}_3$,
and $\Omega_3=\hat{\bf e}_2\cdot\partial_s\hat{\bf e}_1$,
it is straightforward to express (\ref{eq1}) in terms of ${\bf r}$
and $\theta$.

In our dynamic simulation, the filament is modelled as a chain
of $N+1$ connected spheres of diameter $a$. 
Each bead is specified by its position ${\bf r}_j$ and the twist
angle with respect to the laboratory frame, $\theta_j$,
from which the strain vector $\vecOm_j$ is calculated at each sphere point.
The unit tangent is thus given by ${\bf t}_j={\bf u}_j/|{\bf u}_j|$, where
${\bf u}_j={\bf r}_{j+1}-{\bf r}_j$ is  the bond vector.
The simplest symmetric discretization of $\Omega_3$ is
by $\Omega_{3,j}=[\frac{1}{2}({\bf e}_2\cdot\partial_s{\bf e}_1
-{\bf e}_1\cdot\partial_s{\bf e}_2)]_j=f_j\sin(\theta_j-\theta_{j-1})
+g_j\cos(\theta_j-\theta_{j-1})$, where $f$ and $g$ are
functions of $\{{\bf r}_j\}$ only.
The total elastic energy $E$ involves the stretching 
contribution: 
$E_{tot}=E[{\bf r},\theta]+\sum_{j=1}^{N-1}
K/(2a)(|{\bf u}_j|-a)^2$.
For an isotropic rod, the stretching modulus $K$ is given by
 $K=16 A  / a^2$.
Writing the angular velocity as  
$\partial_t\phi=\partial_t\theta+{\bf b}\cdot\partial_t{\bf n}$ 
and using $\nabla_{{\bf r}_j}\theta=-{\bf b}\cdot\nabla_{{\bf r}_j}
{\bf n}$,
we arrive at the coupled Langevin equations
\begin{equation}
 \dot{\bf r}_i(t) = \sum_{j=1}^{N+1}\vecmu_{ij}({\bf r}_{ij})
	\left[-\left.\nabla_{{\bf r}_j}E_{tot}\right|_{\theta}
	+\sum_{k=1}^{N+1}\left.\nabla_{{\theta}_k}E_{tot}\right|_{{\bf r}}
	{\bf b}_k\cdot\nabla_{{\bf r}_j}{\bf n}_k\right]+\vecxi_i(t),
 \label{eq-5a}
\end{equation}
\begin{equation}
 \dot\theta_i(t) = -{\bf b}_i\cdot\dot{\bf n}_i
	-\mu_r\nabla_{\theta_j}E_{tot}+\Xi_i(t).
 \label{eq-5b}
\end{equation}
Hydrodynamic interactions between two spheres $i$ and $j$ are
included via the Rotne-Prager mobility tensor
$\vecmu_{ij}({\bf r}_{ij})=1/(8\pi\eta r_{ij})[{\bf 1}
	+\hat{\bf r}_{ij}\hat{\bf r}_{ij}
	+a^2/(2r_{ij}^2)({\bf 1}/{3}
	-\hat{\bf r}_{ij}\hat{\bf r}_{ij})]$,
where $\hat{{\bf r}}_{ij}={\bf r}_{ij}/r_{ij}$ and 
$\eta$ the solvent viscosity~\cite{ermak}.
For the translational and rotational  self-mobilities of 
the spherical monomers with diameter $a$ we use
$\vecmu_{ii}={\bf 1}/(3\pi\eta a)\equiv\mu_0{\bf 1}$
and  $\mu_r\approx 1/\pi\eta a^3$~\cite{happel}.
The hydrodynamic coupling of rotation and translation
between two spheres, the so-called {\it rotlet} effect~\cite{happel},
decays fast in space ($\sim 1/r^2$) and is
neglected, similar to previous  studies~\cite{wolgemuth}.
Rotlet corrections are  examined  separately~\cite{preparation}.
The vectorial random displacements $\vecxi(t)$ and $\Xi(t)$ model
the coupling to a heat bath and obey the fluctuation-dissipation
relations 
$\bra\vecxi_i(t)\vecxi_j(t')\ket=2k_BT\vecmu_{ij}\delta(t-t')$,
$\bra\Xi_i(t)\Xi_j(t')\ket=2k_BT\mu_r\delta_{ij}\delta(t-t')$
and $\bra\Xi_i(t)\vecxi_j(t')\ket={\bf 0}$, which are numerically 
implemented by a Cholesky factorization~\cite{ermak}.

\begin{figure}
\onefigure[height=13.4cm]{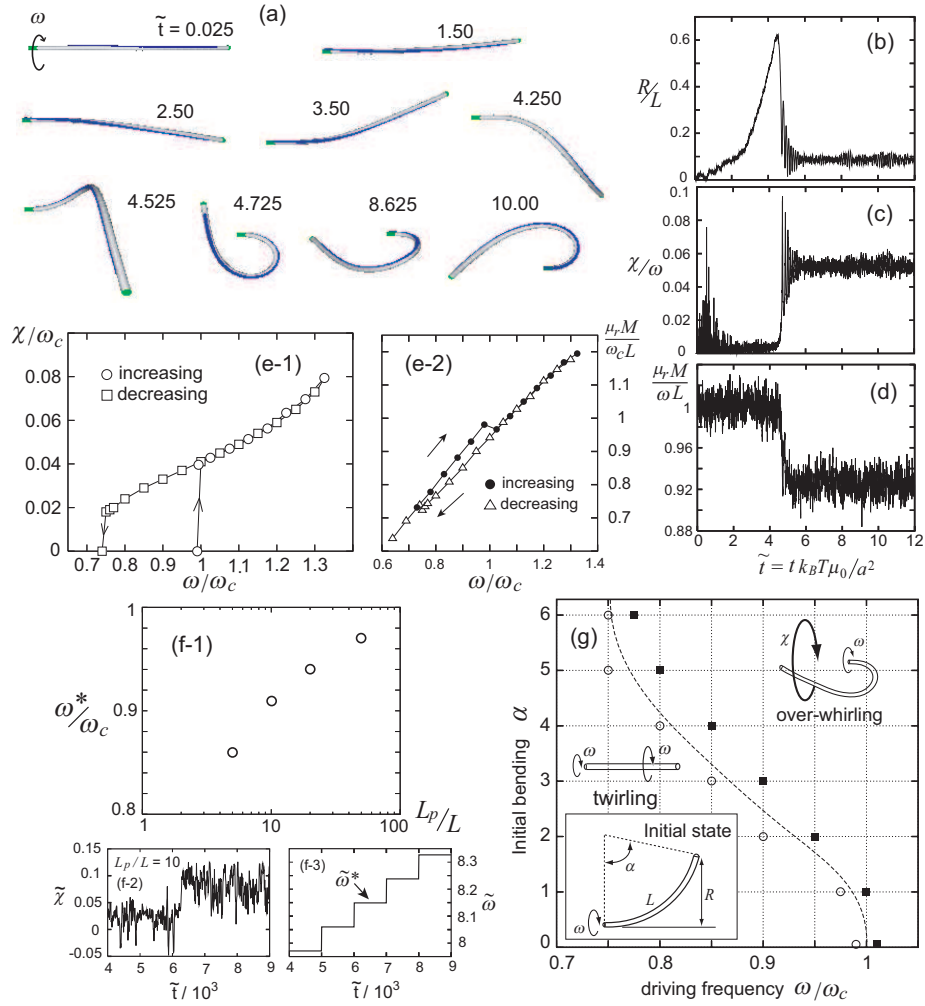}
\caption{(a) Sequence of snapshots of a rotating filament of length
$N=30$ and  stiffness $L_p/L=10^3$ for $\omega/\omega_c=1.20$.
Corresponding time evolutions of (b) the end-point radius $R$, 
(c) the crankshafting frequency $\chi$ and (d) the torque applied at
the driving end $M$.
(e-1) Hysteresis loop of $\chi$ for a $N=30$ and $L_p/L=10^3$ rod.
(e-2) same for $M$.
(f-1) Observed transition frequency $\omega^{\ast}$ for varying 
stiffness $L_p/L$ for $N=30$ rod.
(f-2 and f-3) Corresponding time series of $\chi$ and $\omega$ close to the
transition for $L_p/L=10$.
(g) Zero-temperature stability diagram as a function of  initial bending angle $\alpha$
and driving frequency $\omega$ for a $N=30$ rod.
All data are for an immobile filament.}
\label{f.1}
\end{figure}

For the numerical integrations we discretize Eqs. (\ref{eq-5a}) and
(\ref{eq-5b}) with a time step $\Delta$ and rescale all lengths,
times and energies, leading to the dimensionless parameters
$\tilde{\Delta}=\Delta k_BT\mu_0/a^2$, $\tilde{\mu}_r=\mu_ra^2/\mu_0
=3$, $L_p/L=A/(aNk_BT)$ and $\tilde{\omega}=\omega a^2/(\mu_0k_BT)$.
We set the twist-bend rigidity ratio to $C/A=2$.
This choice,  which corresponds  to a negative Poisson ratio
 $\sigma = A/C-1$~\cite{landau}, is
 relevant for typical biopolymers~\cite{vologodskii,tsuda}.
We also studied the $C/A=1$ case, with no qualitative difference~\cite{preparation}.
For sufficient numerical accuracy we choose time steps in the range
$\tilde{\Delta}=10^{-4}$-$10^{-7}$.
Output values are calculated every $10^3$-$10^4$ steps, total 
simulation times are $10^{6}$-$10^{8}$ steps.
Clamped boundary conditions at the forced end,
i.e. $\partial_s{\bf r}(0)=0$,  are realized by fixing the first two monomers
in space  by applying virtual forces,
which also act (via the mobility tensor) on the rest of the filament.
A mobile filament is obtained by switching off the
virtual forces along the axis  (chosen as the $\hat{\bf x}$ direction).
The rotational driving at the base
imposes $\dot{\theta}_1(t)=\omega$.
Force- and torque-free boundary conditions are adopted for the other end.
As initial shape of the filament we take circular sections
specified by the angle $\alpha$; see the inset of fig.~\ref{f.1} (g).
The straight rod configuration thus corresponds to $\alpha=0$.
The number of beads is in the range  $L/a=N=20-40$.

At low rotational frequency, $\omega < \omega_c$, 
the rod is twisted but remains straight and 
the torque at the base balances  the total rotational drag,
$\mu_r^{-1}\omega L \sim C\Omega_3(0)$. On the scaling level,
the rod buckles when the twisting torque $C\Omega_3(0)$ becomes 
comparable to the bending torque, $A/L$, 
giving a critical frequency 
$\omega_c \sim \mu_rA/L^2$ independent of the twist rigidity $C$.
The asymptotically exact linear analysis based on slender-body hydrodynamics 
predicts a critical frequency~\cite{wolgemuth}
\begin{equation}
\omega_c\cong 8.9\mu_rA/L^2=8.9\mu_rk_BT (L_p/L)^2 /L_p.
\label{eq-crit}
\end{equation}
The time evolution of the rod shape for $N=30$, persistence length $L_p/L=10^3$ 
for $\omega/\omega_c=1.20$ 
(starting with a straight shape $\alpha=0$) is shown in fig.~\ref{f.1}a.
Twirling is unstable against thermal disturbance and the filament
buckles to relieve twist,  leading to  crankshafting motion.
As will be discussed in more detail below,
the radial distance of the free end from the rotational axis,
called $R$,  initially increases
exponentially in time until the filament bends over (see  fig.~\ref{f.1}b)
and the steady overwhirling shape is reached.
Overwhirling is a combination of  rigid-body-like 
rotation with  frequency $\chi$ and  axial spinning.
The crankshafting frequency $\chi$ increases drastically across the 
kinetic transformation state, 
while the torque applied at the base, $M$,  shows a small but steep 
decrease  (see  fig.~\ref{f.1}c-d); in other words,
the transition to the whirling state can be viewed as a way
to reduce the dissipated power.
In figs.~\ref{f.1}e-1 and e-2 the steady state values of $\chi$ 
and $M$ are plotted against
$\omega/\omega_c$ across the transition for a rather stiff rod with $L_p/L=10^3$.
A pronounced hysteresis is revealed, suggesting that the observed transition
is strongly discontinuous and the final state depends not only on the driving
frequency $\omega$ but also on the initial shape of the filament (described
by $\alpha$).
This stands in vivid contrast to the analytics 
predicting a continuous transformation\cite{wolgemuth}.
The bifurcation frequencies depend on the rate with which 
the frequency is changed, but only for very flexible rods 
does the hysteresis disappear.
We construct the stability diagram shown in fig.~\ref{f.1} (g)
 in the $(\omega,\alpha)$ plane
at zero temperature (i.e.  the deterministic noise-less limit where
$L_p/L = \infty$)
by following the time evolution of a rod with an initial circular
shape with bending angle $\alpha$ and the stationary twist profile.
Twirling is the final  stationary state for a filament with an initial 
shape bending angle below the separatrics,
while  above the separatrics the overwhirling state is obtained.
The separatrics terminates for  $\alpha=0$ at 
$\omega=\omega_c$ as given by eq. (\ref{eq-crit})~\cite{wolgemuth},
indicating that $\omega_c$ is
an instability point and  actually the correct critical frequency 
at zero-temperature above which the filament is unstable against 
 infinitesimal disturbance. 
By construction, the stability diagram does not 
depend on the stiffness of the filament.
The cause of the discrepancy to
the previous simulation~\cite{lim} at zero temperatures
which yielded a much lower transition frequency is unclear.

At finite temperature (i.e. finite $L_p/L$), 
the overwhirling transition is observed below $\omega_c$:
In fig.~\ref{f.1} (f-1) the  transition frequency $\omega^{\ast}$ is
plotted as a function of $L_p/L$, obtained from simulations where
the driving  frequency $\omega$ is increased in steps 
of size $\Delta \omega = \omega_c/ 100$ every
 $10^7$ simulation time steps.
The time-dependent behavior of $\chi$ and $\omega$ 
across the transition is plotted
in fig.~\ref{f.1} (f-2) and (f-3), respectively, for $L_p/L=10$,
from which $\omega^{\ast}$ is determined.
An even slower frequency increase rate gives a slightly 
lower $\omega^{\ast}$, so the results in fig.~\ref{f.1} (f-1)
correspond to an upper limit of the transition in the stationary limit.
A simple scaling arguments explains the observed trend:
Balancing the bending energy $E\sim A \alpha^2/ (2 L)$ with 
$k_BT/2$, one obtains a spontaneous  mean squared
bending angle  $ \langle \alpha^2 \rangle 
\sim L/L_p$, which is larger
for smaller $L_p$ and thus leads to
 reduced stability (see the stability diagram).
Below $L_p/L\sim5$, the filament switches
frequently between the two states, 
and the transition is washed out.
The dependence of the twirling-overwhirling transition on the
persistence length could be experimentally tested with 
actin or tubulin filaments of varying length.

\begin{figure}
\onefigure[height=8.7cm]{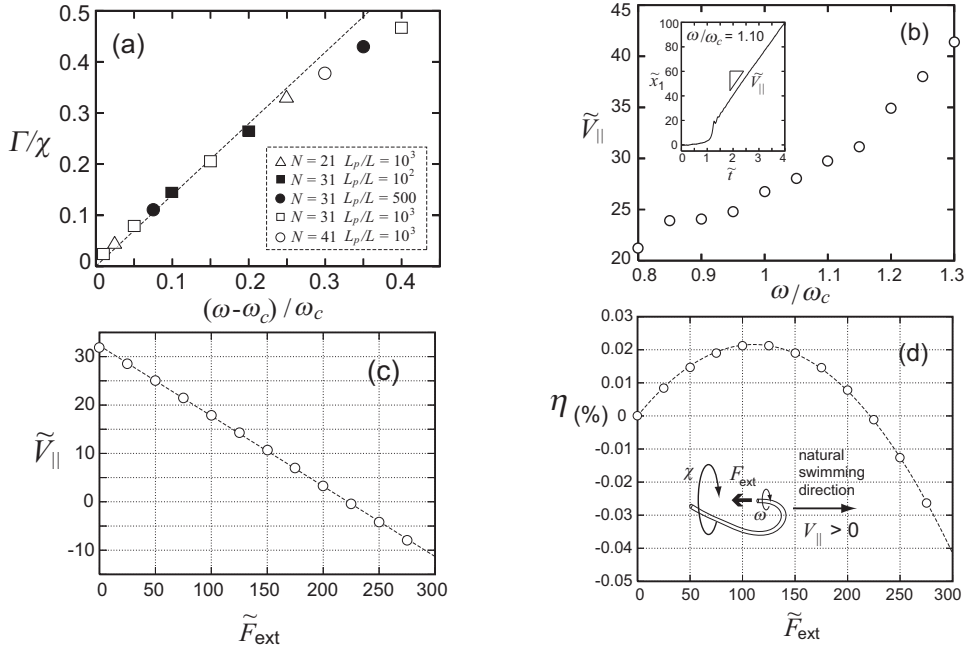}
\caption{(a) Growth rate of end radius
$\Gamma$ divided by the crankshafting frequency
$\chi$ as a function of the supercriticality parameter 
$\varepsilon=(\omega-\omega_c)/\omega_c$ for rods with
various length and stiffness obtained in the noise-less limit. 
(b) Rescaled propulsion 
velocity along the rotation
axis $\tilde{V}_{\parallel}=V_{\parallel}a/\mu_0k_BT$ in the overwhirling state,
plotted against $\omega/\omega_c$ (Inset: actual
time evolution of the base position $\tilde{x} _1$ across the shape 
transition for $\varepsilon=0.1$). 
(c) $V_{\parallel}$ versus the rescaled external force 
$\tilde{F}_{ext}=a F_{ext}/k_BT$ for $\varepsilon=0.1$.
(d) Efficiency $\eta$ as a function of $\tilde{F}_{ext}$ with the 
parabolic fit (dashed line).
All data in (b)-(d) are for $N=30$ and $L_p/L=10^3$ (i.e. at finite temperature).}
\label{f.2}
\end{figure}

For a stiff rod and short times, 
we expect universal shape
dynamics for frequencies slightly above $\omega_c$.
Consider a small amplitude whirling motion ($R/L\ll 1$) 
at crankshafting frequency $\chi$ for supercritical frequency
$\varepsilon\equiv(\omega-\omega_c)/\omega_c>0$.
As we will show, the small amplitude whirling is
unstable for any $\varepsilon$, 
but lasts for a  long time for small enough $\varepsilon$ due to critical slowing down. 
The rod shape is approximated as 
${\bf r} (s) \approx s\hat{\bf x}+{\bf r}_{\perp}(s)$, 
and  $\dot{\bf r}(s)=\chi\hat{\bf x}\times{\bf r}_{\perp}(s)$ for crankshafting.
Since $\chi$ grows much more slowly in time than the end-point radius $R$,
it can be estimated by assuming that the transverse drag force per length,
$\chi |{\bf r}_{\perp}| /\mu_0$, is roughly equal to the bending
force per length, $A|{\bf r}_{\perp}|/L^4$, giving 
$\chi\sim A \mu_0 / L^4\sim (a/2L)^2\omega_c$.
After a short time transient, the twist density $\Omega_3$ has built up the stationary
linear profile, $\Omega_3=\omega/\mu_rC(s-L)$,  
and thereby reached $\dot{\Omega}_3=0$.
The difference in rotational velocities about the local tangents at $s=0$
and $s=L$, $\Delta\omega$,  satisfies 
$-\Delta\omega = \chi\left[1-\hat{\bf x}\cdot{\bf t}(L)\right]$
(for a derivation  see ref.~\cite{wolgemuth}).
Considering the local balances of viscous and elastic twisting torques
$\mu_r^{-1}\omega L\sim C\Omega_3$ at $s=0$ and $s=L$, we see that
the net torque $\Delta M\sim \mu_r^{-1}\Delta\omega L$ must contribute to the
whirling motion of the filament.
On the other hand, the elastic energy $E$ of the rod with the amplitude $R$
is estimated to be $E\sim 2AR^2/L^3$, while the power dissipation $P_d$
by the rigid-body like rotation is $P_d\sim v^2L/3\mu_0=\chi^2R^2L/3\mu_0$,
where $v=\chi R$ is the rotational velocity.
The power balance condition, according to which
the sum of the change of the elastic energy per unit time
and the hydrodynamic rigid-body dissipation balances the net torque 
injected into the rod per unit time, implies $\dot{E}+P_d\sim \omega \Delta M$.
Note that the remaining power input $\omega(M-\Delta M)$ is independently
balanced with the dissipation by axial rotation.
Approximating $1-\hat{\bf x}\cdot{\bf t}(L) \approx 2R^2/L^2$
we finally arrive at
$\dot{R}(t)\sim\varepsilon\chi R(t)$.
The end point radius $R(t)$ thus evolves exponentially in time 
for $\varepsilon>0$, and 
the growth rate $\Gamma(\omega)=d(\log R)/dt$ satisfies the linear relation
$\Gamma/\chi\sim (\omega-\omega_c)/\omega_c$.
In fig.~\ref{f.2} (a), the numerically obtained $\Gamma/\chi$, which is almost
constant while $R/L \ll 1$, is plotted as a function of 
the supercriticality 
$\varepsilon=(\omega-\omega_c)/\omega_c$ for filaments of various length and
stiffness. The thermal noise 
is switched off for improved data accuracy.
The data scale linearly for small $\varepsilon$ in agreement with the prediction 
(with a numerical prefactor of the order of unity), 
verifying the validity of the arguments and approximations presented above.  
For larger $\varepsilon$, nonlinearity comes in, leading to deviations
from the simple linear law.
For the numerical calculations the presence
of this critical slowing down can be a problem
because stationary states are only obtained after a long waiting time.

In the case when the filament is allowed to move along
the axis of rotation, a finite average velocity is observed in the
overwhirling regime: the filament is propelled due to the 
formation of a helical shape which breaks the time-reversal 
symmetry of the problem.
Accordingly, a jump in the propulsion velocity along 
the rotation axis $V_{\parallel}$ is observed across the overwhirling transition, 
resulting in a drastically amplified forward thrust of the filament
(see the inset of fig.~\ref{f.2} (b)). 
The propulsion velocity $V_{\parallel}$ in the overwhirling regime
is plotted in fig.~\ref{f.2} (b) as a function of $\omega/\omega_c$.
Note that the overwhirling state is metastable for a wide
range of frequencies below $\omega^*$ and
thus the velocity is non-zero. The
non-monotonic increase of $V_{\parallel}$ with $\omega$
implies a complicated shape
change of the rod as a function of  $\omega$.
To study the swimming efficiency of this self-propelling object,
we apply an external force $F_{ext}$ at the filament base.
The power converter efficiency is the ratio of the propulsive
power output and the rotary power input,
$\eta=F_{ext}V_{\parallel}/(M\omega)$, where a positive $F_{ext}$ 
is defined such as to work against the 
natural propulsion direction of the filament.
The rotational and translational mobilities of the whole filament 
are defined by
$\omega=\mu_{rr}M+\mu_{rt}F_{ext}$ and 
$V_{\parallel}=\mu_{tr}M+\mu_{tt}F_{ext}$~\cite{purcell}. 
Although for perfectly stiff propellers the mobility matrix is constant
and symmetrical (i.e., $\mu_{rt}=\mu_{tr}$),
due to the flexibility of the filament,
 the symmetry here is broken
and the mobilities depend 
on the external  torque $M$ and force $F_{ext}$ in a complicated non-linear manner.
In the simulation, a substantial forward thrust is 
observed independent of 
the rotation direction, which suggests that $\mu_{tr}$
changes its sign when the torque $N$ is reversed,
similar to results obtained for a rod with vanishing
twist rigidity that rotates 
on the surface of a cone\cite{manghi}.
$V_{\parallel}$ varies linearly with ${F}_{ext}$ as shown in
fig.~\ref{f.2} (c), which means that $\mu_{tt}$ is independent of 
the external force ${F}_{ext}$.
Neglecting $\mu_{rt}$, which is actually
vanishingly small, while keeping $\mu_{tr}$, we obtain 
$\eta=(\mu_{rr}\mu_{tt}F_{ext}^2+\mu_{tr}\omega F_{ext})/\omega^2$,
indicating that the efficiency becomes parabolic as a function of the
external force. 
Figure~\ref{f.2} (d) shows the numerical data of $\eta$ with the parabolic fit
for $\varepsilon=0.1$, which describes the date quite nicely.
The highest efficiency is only of the
order of $0.01\%$, which means that
this self-propelling filament is  quite inefficient compared to other
known examples involving helices
and other chiral objects~\cite{manghi}, since most of the
input power is dissipated by  the axial spinning.

In summary, the elastohydrodynamics of a spinning filament in viscous fluid
is studied by simulations including full coupling 
between elastic, thermal and hydrodynamic effects.
Quantitative agreement with the analytically predicted critical
frequency is obtained\cite{wolgemuth} in the absence of thermal fluctuations,
in contrast to numerical simulations of the same problem\cite{lim}.
We give evidence for the discontinuous nature of the twirling-overwhirling transition,
in qualitative agreement with previous numerical works at zero temperature\cite{lim}
but in contradiction to the analytical theory which might
have to do with the neglect of non-linear effects\cite{wolgemuth}.
Thermal fluctuations play a significant role in the transition
behavior of a filament and lead to a decrease of the 
critical frequency in a range of filament stiffnesses $L_p/L$  that 
is relevant to biopolymers.
The parametrization proposed in this paper is entirely based upon a
local-form description and advantageous for numerical implementations.
Our dynamic simulation method is 
easily generalizable to other geometries  such as rings or helices.
A number of intriguing problems have to do with
pulling on flexible nanospring~\cite{smith} or buckling of twisted 
rods~\cite{goriely,vologodskii}.
Of particular biological interest is a spinning helical filament.
Taking, for example, a diameter value of $a\approx 20$ nm and 
a typical stiffness 
$L_p/L\approx10^{3}$ of a bacterial flagellum at physiological 
conditions~\cite{turner},
we find $\omega_c \sim 10^3$ s$^{-1}$ in water 
with the viscosity $\eta\sim 10^{-3}$ Pa$\cdot$s, which
is easily achievable in laboratory experiments.
Details on the interplay of thermal fluctuations
and shape instabilities will be reported elsewhere~\cite{preparation}.

We thank M. Manghi for valuable discussions
and the program for  Research Abroad of the Japan Society for
the Promotion of Science (JSPS) and
 the German Science Foundation (DFG, SPP1164) for financial support.


\begin{thebibliography}{99}

\bibitem{goriely}
 \Name{Goriely A. \and Tabor M.}
 \REVIEW{Nonlinear Dynamics}{21}{2000}{101} and references therein.

\bibitem{wiggins}
 \Name{Wiggins C. H.}
 \REVIEW{Math. Meth. Appl. Sci.}{24}{2001}{1325}.
 
\bibitem{kaes}
 \Name{K\"{a}s J., Strey H., B\"{a}rmann M. \and Sackmann E.}
 \REVIEW{Europhys. Lett.}{21}{1993}{865}.
 
\bibitem{landau}
 \Name{Landau L. D. \and Lifshitz E. M.}
  \Book{Theory of Elasticity}
  \Publ{Pergamon Press, Oxford}
  \Year{1980}.
  
\bibitem{wolgemuth}
 \Name{Wolgemuth C. W., Powers T. R., \and Goldstein R. E.}
 \REVIEW{Phys. Rev. Lett.}{84}{2000}{1623}.

\bibitem{lim}
 \Name{Lim S. \and Peskin C. S.}
 \REVIEW{SIAM J. Sci. Comput.}{25}{2004}{2066}.

\bibitem{goldstein}
 \Name{Goldstein R. E., Powers T. R., \and Wiggins C. H.}
 \REVIEW{Phys. Rev. Lett.}{80}{1998}{5232}.

\bibitem{allison}
 \Name{Allison S., Austin R., \and Hogan M.}
 \REVIEW{J. Chem. Phys.}{90}{1989}{3843}.

\bibitem{ermak}
 \Name{Ermak D. L. \and McCammon J. A.}
 \REVIEW{J. Chem. Phys.}{69}{1978}{1352}.

\bibitem{happel}
 \Name{Happel J. \and Brenner H.}
 \Book{Low Reynolds Number Hydrodynamics}
 \Publ{Noordhoff, Leyden}
 \Year{1973}.
  
\bibitem{preparation}
 \Name{Wada H. \and Netz R. R.}
 in preparation.

\bibitem{vologodskii}
 \Name{Vologodskii A. V., \and Marko J. F.}
 \REVIEW{Biophys. J.}{73}{1997}{123}.

\bibitem{tsuda}
 \Name{Tsuda Y., Yashutake H., Ishijima A., \and Yanagida T.}
 \REVIEW{Proc. Natl. Acad. Sci. U.S.A.}{93}{1996}{12937}.

\bibitem{purcell}
 \Name{Purcell E. M.}
 \REVIEW{Proc. Natl. Acad. Sci. U.S.A.}{94}{1997}{11307}.

\bibitem{manghi}
 \Name{Manghi M., Schlagberger X., \and Netz R. R.}
 \REVIEW{Phys. Rev. Lett.}{96}{2006}{068101}.

\bibitem{smith}
 \Name{Smith B., Zastavker Y. V. \and Benedek G. B.}
 \REVIEW{Phys. Rev. Lett.}{87}{2001}{278101}.

\bibitem{turner}
 \Name{Turner L., Ryu W. S., \and Berg H. C.}
 \REVIEW{J. Bacteriol.}{182}{2000}{2793}.

\end{thebibliography}
\end{document}